\documentclass[journal]{IEEEtran}
\usepackage[utf8]{inputenc}
\usepackage{graphicx} 
\usepackage{amsmath}
\usepackage[linesnumbered,ruled,vlined]{algorithm2e}
\usepackage{float}
\usepackage{subcaption}
\usepackage{enumerate}
\usepackage{color}
\usepackage[margin=1in]{geometry}
\usepackage{cleveref}
\usepackage{booktabs}
\usepackage{authblk}
\usepackage{cleveref}
\usepackage{acronym}
\usepackage{amsmath}
\usepackage{makecell}

\crefname{figure}{Fig.}{Figs.}
\Crefname{figure}{Fig.}{Figs.}

\title{Dual-Satellite Doppler Accuracy Prediction and Geometry Selection for Sparse LEO Signals of Opportunity}
 \author{Qi Liu, Marc Fernández-Temprado, Antoni Reus-Bergas, Shuguo Pan, Wang Gao, Gonzalo Seco-Granados, José A. López-Salcedo
 \thanks{This work has been partly supported by the Catalan Government
in the framework of the NewSpace Strategy of Catalonia and by the
Spanish Agency of Research (AEI) under grant PID2023-152820OB-I00
funded by MICIU/AEI/10.13039/501100011033 and by ERDF/EU.}}

\date{February 2026}

\begin{document}

\maketitle
\begin{abstract}
Low Earth Orbit (LEO) satellites have recently attracted increasing attention as complementary sources for positioning, navigation, and timing in GNSS-challenged environments. In sparse LEO signals-of-opportunity scenarios, Doppler positioning may rely on only one or two satellite passes within a limited observation period, making the achievable accuracy strongly dependent on pass geometry. This paper investigates dual-satellite LEO Doppler accuracy prediction, geometry selection, and temporal availability. A single-pass Doppler accuracy representation, derived from the Doppler-based Dilution of Precision (DDOP) framework, is first validated using real Iridium Doppler measurements by comparing predicted error ellipses with empirical positioning error distributions. An information-domain dual-pass accuracy model is then developed by fusing the effective position information from two satellite passes while accounting for pass-specific clock drift and timing correction parameters. Based on this model, an analytical relationship is derived between the error-ellipse intersection angle and the fused dual-satellite positioning accuracy, where the intersection angle is defined as the angle between the major axes of two single-pass predicted error ellipses. Controlled simulations confirm that the proposed closed-form angle–accuracy expression agrees well with sample-based positioning results. Long-term ORBCOMM observations are further used to evaluate the practical significance of the proposed geometry model. The results show that an intersection angle of approximately $20^\circ$ is required to achieve a theoretical positioning accuracy of about 50 m, and that a complementary second pass obtained within about 30 min can generally support sub-50 m theoretical accuracy. These findings provide practical guidance for geometry-aware satellite-pair selection, observation scheduling, and accuracy availability assessment in sparse LEO Doppler positioning.
\end{abstract}


\section{Introduction}
Conventional Global Navigation Satellite System (GNSS) has become the primary infrastructure for positioning, navigation, and timing (PNT) system, but its performance can be severely degraded in challenging environments such as urban canyons, indoor-like areas, and interference-prone scenarios \cite{zhu2018gnss,zangenehnejad2021gnss}. As a complementary source of PNT information, Low Earth Orbit (LEO) satellites have recently attracted increasing attention due to their lower orbital altitude, stronger received signal power, and rapidly varying observation geometry \cite{liu2026innovative,eissfeller2024comparative}. In particular, the fast motion of LEO satellites produces pronounced Doppler variations, which can be exploited for positioning even when dedicated navigation messages are unavailable \cite{kozhaya2023positioning}. Therefore, Doppler-based positioning using LEO signals of opportunity provides a promising alternative for GNSS-challenged environments \cite{jiang2022leo}.

Nevertheless, LEO Doppler positioning in signals-of-opportunity scenarios is fundamentally limited by sparse satellite visibility and geometry \cite{liu2026geometric, psiaki2021navigation}. Unlike conventional GNSS positioning, where multiple satellites are typically observed simultaneously, a receiver may only have access to one or a few useful LEO satellites over a short observation period. Previous studies have shown that single pass LEO Doppler positioning is strongly geometry-dependent and often highly anisotropic: the along-track direction is well constrained by Doppler variation along the pass, while the orthogonal cross-track direction can remain weakly observable. This behavior can be represented by a predicted error ellipse, which provides a compact description of both the magnitude and the dominant direction of single pass positioning uncertainty. However, such a single pass representation does not directly answer how two passes should be contained, or what relative geometry makes a dual-pass configuration effective.

A key question is therefore how to predict the accuracy of a dual-pass Doppler solution from the accuracy characteristics of two individual passes, and how to determine whether the two passes provide complementary rather than redundant geometric constraints\cite{farhangian2020multi}. In a dual-satellite case, the two passes can reduce the directional ambiguity of each other only when their geometries are sufficiently complementary. If the two satellite passes provide similar geometric constraints, the fused solution may still suffer from poor observability in certain directions, even though additional Doppler measurements are introduced. Therefore, an effective dual-satellite configuration should be determined not only by satellite availability, but also by the relative geometry between the two single-satellite Doppler passes. In this paper, this complementarity is quantified by the intersection angle of ellipse error, defined as the angle between the major axes of two single-pass predicted error ellipses.

Existing studies have demonstrated that using multiple LEO satellites or multiple constellations can improve Doppler positioning performance by increasing measurement diversity and observation geometry \cite{kassas2023navigation}. However, most existing works mainly evaluate the benefit of multiple satellites through positioning results \cite{allahvirdi2025doppler}, receiver implementation \cite{ai2025multi}, or constellation-level geometry metrics \cite{zhang2025efficient,ccelikbilek2025optimization}. The underlying mechanism by which two individual satellite passes complement each other in terms of theoretical accuracy remains insufficiently characterized. In particular, even if the uncertainty of each individual pass is represented by an error ellipse, it remains unclear how the relative orientation of two such ellipses determines the fused dual-pass accuracy, and how this relationship can be used for practical satellite-pair selection and observation scheduling.

Based on the above motivations, the main contributions of this paper are summarized as follows. Firstly, a single-pass Doppler accuracy representation is validated using real LEO measurements. Unlike the previous simulation-based analysis, this paper uses real Iridium observations to examine whether predicted error ellipses are consistent with empirical positioning error distributions. This validation provides the experimental basis for using single satellite error ellipses as the building block for dual-satellite accuracy prediction.

Secondly, an information-domain dual-satellite accuracy prediction model is developed. The proposed model combines the effective geometric information provided by two individual satellite passes while accounting for satellite-specific clock drift and timing offset parameters. By eliminating these nuisance parameters from the information matrix, the equivalent position information matrix of the dual-satellite system is obtained, enabling theoretical accuracy prediction for the fused solution.

Thirdly, an intersection-angle-based geometry model is derived to reveal how the relative orientation of two single-pass error ellipses affects the fused positioning accuracy. The intersection angle is defined as the angle between the major axes of the two predicted error ellipses. Based on the individual-pass accuracy representations, the proposed model establishes an analytical relationship between this angle and the dual-pass positioning accuracy.

Finally, long-term ORBCOMM measurements are used to translate the proposed angle–accuracy relationship into practical satellite selection and observation scheduling guidelines. Based on the ten-day real observation dataset, the minimum intersection angle required for a target accuracy is statistically identified. The analysis further connects geometric complementarity with temporal availability: after one usable pass has been observed, the receiver can estimate how long it may need to wait for a sufficiently complementary second pass, or what accuracy can be expected after a given waiting period. This provides a practical basis for geometry-aware satellite-pair selection, positioning availability assessment, and observation scheduling in sparse LEO signals-of-opportunity PNT scenarios.

\section{Methodology}
This section develops the proposed dual-satellite accuracy prediction and geometry-selection framework. First, a compact Doppler observation model is introduced to define the measurement and state notation. Then, a dual-pass Doppler positioning model is formulated by enforcing a common receiver position while allowing pass-specific timing-related parameters. Based on this model, the effective geometric information matrix is derived by eliminating nuisance parameters through the Schur complement. Finally, an analytical relationship is established between the error-ellipse intersection angle and the fused dual-pass positioning accuracy.
\subsection{LEO Satellite Doppler Shift Signal Model}
For each satellite pass, the Doppler observation is modeled as a nonlinear function of the receiver position and pass-specific timing-related parameters. The receiver is assumed to remain stationary within the observation window, so that a common receiver position can be used for all Doppler measurements in the same pass. The pass-specific parameters include the receiver clock drift and a timing correction term used to account for the time-domain uncertainty in the satellite ephemeris evaluation. The Doppler observation at epoch $t$ is expressed in range-rate units as:

\begin{equation}
z_s(t)=h_s(t;\boldsymbol{\theta}_s)+\epsilon_s(t).\label{eqDopShift}
\end{equation}
where
\begin{equation}
    h_s(t;\boldsymbol{\theta}_s) = -\mathbf{v}_s(t - \delta_{c,s})^T \frac{\mathbf{r} - \mathbf{r}_s(t - \delta_{c,s})}{\|\mathbf{r}-\mathbf{r}_s(t-\delta_{c,s})\|} + c\dot\delta_{d,s}\label{eqDopSigModel}
\end{equation}
Here, $\mathbf{r}$ is the receiver position, $\mathbf{r}_s(\cdot)$ and $\mathbf{v}_s(\cdot)$ are the satellite position and velocity, $c$ is the speed of light, $\dot{\delta}_{d,s}$ is the receiver clock drift, and $\delta_{c,s}$ is the timing correction parameter. The unknown vector is defined as
\begin{equation}
\boldsymbol{\theta}_s=
[\mathbf{r}^{T},\boldsymbol{\kappa}_s^{T}]^{T},
\quad
\boldsymbol{\kappa}_s=
[\dot{\delta}_{d,s},\delta_{c,s}]^{T}.
\label{eqStateSinglePass}
\end{equation}

For a pass containing $M$ Doppler measurements, the observations are stacked as
\begin{equation}
\mathbf{z}_s=
[z_s(t_1^s),z_s(t_2^s),\ldots,z_s(t_M^s)]^T .
\label{eqStackedZ}
\end{equation}
The stacked observation model is then written as
\begin{equation}
\mathbf{z}_s=
\mathbf{h}_s(\boldsymbol{\theta}_s)
+\boldsymbol{\epsilon}_s .
\label{eqDefZ}
\end{equation}
where $\mathbf{h}_s(\boldsymbol{\theta}_s)$ and $\boldsymbol{\epsilon}_s$ denote the stacked nonlinear Doppler function and measurement noise vector, respectively. This model defines the notation used in the following dual-satellite positioning and geometry analysis.

\subsection{Dual-Satellite Doppler Positioning Model}
Based on the Doppler observation model, the measurements from each satellite pass are linearized around a common reference state. The purpose of this subsection is to formulate the dual-pass positioning model, in which the two passes share the same receiver position while keeping their timing-related parameters independent.

For $i$-th satellite pass, the Doppler model is linearized around a common initial state $\boldsymbol{\theta}_{0}$, which is used for all satellite passes.
\begin{equation}
\mathbf{h}_i(\boldsymbol{\theta}_i)
\approx
\mathbf{h}_i(\boldsymbol{\theta}_0)
+
\mathbf{H}_i(\boldsymbol{\theta}_0)
\Delta\boldsymbol{\theta}_i ,
\label{eqLinearModel}
\end{equation}
The corresponding residual vector is
\begin{equation}
\Delta\mathbf{z}_i
=
\mathbf{z}_i-\mathbf{h}_i(\boldsymbol{\theta}_0)
\approx
\mathbf{H}_i(\boldsymbol{\theta}_0)
\Delta\boldsymbol{\theta}_i
+
\boldsymbol{\epsilon}_i .
\label{eqSinglePassResidual}
\end{equation}
Since the receiver position is common to different satellite passes, while the timing-related parameters are pass-specific, the correction vector of the $i$-th pass is partitioned as
\begin{equation}
\Delta\boldsymbol{\theta}_i
=
[
\Delta\mathbf{r}^{T},
\Delta\boldsymbol{\kappa}_i^{T}
]^T .
\label{eqSinglePassPartition}
\end{equation}
Accordingly, the Jacobian matrix is partitioned as
\begin{equation}
\mathbf{H}_i
=
[
\mathbf{A}_i
\quad
\mathbf{B}_i
],
\end{equation}
where $\mathbf{A}_i$ contains the Jacobian columns with respect to the receiver position correction, and $\mathbf{B}_i$ contains the Jacobian columns with respect to the pass-specific timing-related parameters. For compactness, the dependence on the common reference state $\boldsymbol{\theta}_0$ is omitted in the following derivation.

The linearized model of the $i$-th pass can therefore be written in block form as
\begin{equation}
\Delta\mathbf{z}_i
\approx
\mathbf{A}_i\Delta\mathbf{r}
+
\mathbf{B}_i\Delta\boldsymbol{\kappa}_i
+
\boldsymbol{\epsilon}_i .
\label{eqSinglePassBlockModel}
\end{equation}

For dual-satellite Doppler positioning, the two satellite passes are jointly processed by a common receiver position. The joint correction vector is defined as
\begin{equation}
\Delta\boldsymbol{\theta}_{12}
=
[
\Delta\mathbf{r}^{T},
\Delta\boldsymbol{\kappa}_1^{T},
\Delta\boldsymbol{\kappa}_2^{T}
]^T .
\label{eqDualPassState}
\end{equation}
The stacked residual and noise vectors are
\begin{equation}
\Delta\mathbf{z}_{12}
=
[
\Delta\mathbf{z}_1^{T},
\Delta\mathbf{z}_2^{T}
]^T,
\quad
\boldsymbol{\epsilon}_{12}
=
[
\boldsymbol{\epsilon}_1^{T},
\boldsymbol{\epsilon}_2^{T}
]^T .
\label{eqDualPassResidual}
\end{equation}
The dual-satellite linearized model is then given by
\begin{equation}
\Delta\mathbf{z}_{12}
\approx
\mathbf{H}_{12}
\Delta\boldsymbol{\theta}_{12}
+
\boldsymbol{\epsilon}_{12},
\label{eqDualPassModel}
\end{equation}
with
\begin{equation}
\mathbf{H}_{12}
=
\begin{bmatrix}
\mathbf{A}_1 & \mathbf{B}_1 & \mathbf{0}\\
\mathbf{A}_2 & \mathbf{0} & \mathbf{B}_2
\end{bmatrix}.
\label{eqDualPassJacobian}
\end{equation}

This block structure shows that both satellite passes contribute to the same receiver position estimate, while each pass retains its own timing-related nuisance parameters. Therefore, the dual-pass positioning model provides the basis for deriving the effective geometric information matrix in the next subsection.

\subsection{Geometric Information Matrix for Dual-Satellite Accuracy Prediction}
Based on the dual-pass positioning model, this subsection derives the effective geometric information matrix used for theoretical accuracy prediction. The key idea is to eliminate the pass-specific timing-related nuisance parameters and retain only the equivalent information associated with the common receiver position. For compactness, the dependence on the common linearization point $\boldsymbol{\theta}_0$ is omitted in the following derivation. The Jacobian matrices are assumed to be properly normalized before forming the information matrix to handle the different physical units of position and timing-related parameters. Starting from the block model in the previous subsection, the dual-satellite Jacobian and measurement weight matrix are written as

\begin{equation}
\mathbf{H}_{12}
=
\begin{bmatrix}
\mathbf{A}_1 & \mathbf{B}_1 & \mathbf{0}\\
\mathbf{A}_2 & \mathbf{0} & \mathbf{B}_2
\end{bmatrix},
\quad
\mathbf{W}_{12}
=
\begin{bmatrix}
\mathbf{W}_1 & \mathbf{0}\\
\mathbf{0} & \mathbf{W}_2
\end{bmatrix},
\label{eqDualPassHW}
\end{equation}
where $\mathbf{W}_i$ is the measurement weight matrix of the $i$-th pass. In practice, $\mathbf{W}_i$ can be constructed from the Doppler measurement noise variance, with possible elevation-dependent weighting.

Under the WLS formulation, the weighted normal matrix corresponds to the information matrix of the dual-satellite system,
\begin{equation}
\mathbf{N}_{12}
=
\mathbf{H}_{12}^{T}
\mathbf{W}_{12}
\mathbf{H}_{12}.
\label{eqDualPassInfo}
\end{equation}
Substituting the block structure of $\mathbf{H}_{12}$ and $\mathbf{W}_{12}$ into Eq.~(\ref{eqDualPassInfo}) gives
\begin{equation}
\mathbf{N}_{12}
=
\begin{bmatrix}
\mathbf{A}_1^T\mathbf{W}_1\mathbf{A}_1+\mathbf{A}_2^T\mathbf{W}_2\mathbf{A}_2
&
\mathbf{A}_1^T\mathbf{W}_1\mathbf{B}_1
&
\mathbf{A}_2^T\mathbf{W}_2\mathbf{B}_2
\\
\mathbf{B}_1^T\mathbf{W}_1\mathbf{A}_1
&
\mathbf{B}_1^T\mathbf{W}_1\mathbf{B}_1
&
\mathbf{0}
\\
\mathbf{B}_2^T\mathbf{W}_2\mathbf{A}_2
&
\mathbf{0}
&
\mathbf{B}_2^T\mathbf{W}_2\mathbf{B}_2
\end{bmatrix}.
\label{eqDualPassInfoBlock}
\end{equation}

Since the receiver position is the parameter of interest, while the clock drift and timing correction terms are nuisance parameters, $\mathbf{N}_{12}$ is partitioned as
\begin{equation}
\mathbf{N}_{12}
=
\begin{bmatrix}
\mathbf{N}_{pp} & \mathbf{N}_{pt}\\
\mathbf{N}_{tp} & \mathbf{N}_{tt}
\end{bmatrix},
\label{eqInfoPartition}
\end{equation}
where the subscript $p$ denotes the receiver-position parameters and the subscript $t$ denotes the pass-specific timing-related parameters. By eliminating the nuisance parameters through the Schur complement, the equivalent position information matrix is obtained as
\begin{equation}
\mathbf{N}_{pos}
=
\mathbf{N}_{pp}
-
\mathbf{N}_{pt}
\mathbf{N}_{tt}^{-1}
\mathbf{N}_{tp}.
\label{eqSchurComplement}
\end{equation}

Using the block form in Eq.~(\ref{eqDualPassInfoBlock}), the equivalent position information matrix can be expanded as
\begin{equation}
\begin{aligned}
\mathbf{N}_{pos}
&=\\
&\mathbf{A}_1^T\mathbf{W}_1\mathbf{A}_1
-
\mathbf{A}_1^T\mathbf{W}_1\mathbf{B}_1
\left(
\mathbf{B}_1^T\mathbf{W}_1\mathbf{B}_1
\right)^{-1}
\mathbf{B}_1^T\mathbf{W}_1\mathbf{A}_1
\\
&+
\mathbf{A}_2^T\mathbf{W}_2\mathbf{A}_2
-
\mathbf{A}_2^T\mathbf{W}_2\mathbf{B}_2
\left(
\mathbf{B}_2^T\mathbf{W}_2\mathbf{B}_2
\right)^{-1}
\mathbf{B}_2^T\mathbf{W}_2\mathbf{A}_2 .
\end{aligned}
\label{eqNposExpanded}
\end{equation}

The effective geometric information matrix contributed by the $i$-th pass is therefore defined as
\begin{equation}
\mathbf{G}_i
=
\mathbf{A}_i^T\mathbf{W}_i\mathbf{A}_i
-
\mathbf{A}_i^T\mathbf{W}_i\mathbf{B}_i
\left(
\mathbf{B}_i^T\mathbf{W}_i\mathbf{B}_i
\right)^{-1}
\mathbf{B}_i^T\mathbf{W}_i\mathbf{A}_i .
\label{eqEffectiveGeoInfo}
\end{equation}
The matrix $\mathbf{G}_i$ represents the position-related geometric information provided by the $i$-th pass after accounting for its coupling with the pass-specific timing-related nuisance parameters. Therefore, the equivalent position information matrix of the dual-pass system becomes
\begin{equation}
\mathbf{N}_{pos}
=
\mathbf{G}_1+\mathbf{G}_2 .
\label{eqNposSum}
\end{equation}

The theoretical position covariance of the dual-pass solution is then given by
\begin{equation}
\mathbf{P}_{12}
=
\mathbf{N}_{pos}^{-1}
=
(\mathbf{G}_1+\mathbf{G}_2)^{-1}.
\label{eqDualPassCovariance}
\end{equation}
This result shows that the theoretical accuracy of the dual-pass solution is governed by the sum of two effective geometric information matrices. Therefore, the benefit of adding a second pass depends not only on the number or quality of additional Doppler measurements, but also on whether $\mathbf{G}_1$ and $\mathbf{G}_2$ constrain complementary directions. This motivates the intersection-angle-based accuracy analysis developed in the next subsection.

\subsection{Intersection-Angle–Accuracy Relationship}

Building on the effective geometric information matrices derived above, the positioning accuracy of the fused solution depends not only on the individual accuracy contribution of each satellite pass, but also on the geometric complementarity between the two passes. To analyze the effect, this subsection derives the relationship between the intersection angle and the theoretical positioning accuracy based on the two-dimensional horizontal position covariance. Here, the intersection angle is represented by the angle between the major axes of the two single-pass predicted error ellipses.

For the $i$-th satellite pass, let $\mathbf{P}_{i,2D}$ denote its effective two-dimensional position covariance matrix in the local horizontal plane. It can be obtained from the predicted position covariance and represented through eigenvalue decomposition as

\begin{equation}
{\mathbf{P}}_{i,2D}
=
\mathbf{R}(\phi_i)
\begin{bmatrix}
\lambda_{i,\max} & 0 \\
0 & \lambda_{i,\min}
\end{bmatrix}
\mathbf{R}^T(\phi_i) ,
\end{equation}
where $\lambda_{i,\max}$ and $\lambda_{i,\min}$ are the major- and minor-axis variances of the error ellipse, respectively. The angle $\phi_i$denotes the orientation of the major axis of the $i$-th error ellipse with respect to the reference coordinate frame. The corresponding effective 2D position information matrix in the horizontal plane is
\begin{equation}
\mathbf{G}_{i,2D}
=
{\mathbf{P}}_{i,2D}^{-1}
=
\mathbf{R}(\phi_i)
\begin{bmatrix}
\mu_i & 0 \\
0 & \nu_i
\end{bmatrix}
\mathbf{R}(\phi_i)^T .
\end{equation}
where $\mu_i=1/\lambda_{i,\max}$ and $\nu_i=1/\lambda_{i,\min}$. Since $\lambda_{i,\max}\geq\lambda_{i,\min}$, it follows that $\mu_i\leq\nu_i$.
Let $\beta$ denote the error-ellipse intersection angle between the major axes of the two single-pass predicted error ellipses. Without loss of generality, the principal-axis coordinate system of the first error ellipse is selected as the reference frame. Then the two information matrices can be written as
\begin{equation}
\mathbf{G}_{1,2D} =
\begin{bmatrix}
\mu_1 & 0 \\
0 & \nu_1
\end{bmatrix},
\quad
\mathbf{G}_{2,2D} =
\mathbf{R}(\beta)
\begin{bmatrix}
\mu_2 & 0 \\
0 & \nu_2
\end{bmatrix}
\mathbf{R}^T(\beta) ,
\end{equation}

Following the dual-satellite covariance expression in \cref{eqDualPassCovariance}, the fused two-dimensional position covariance is given by

\begin{equation}
    \mathbf{P}_{12,2D} = (\mathbf{G}_{1,2D} +\mathbf{G}_{2,2D})^{-1} = 
\frac{1}{AD-B^2}
\begin{bmatrix}
D & -B \\
-B & A
\end{bmatrix},
\end{equation}
with
\begin{equation}
\begin{aligned}
    A &= \mu_1 + \mu_2 \cos^2\beta + \nu_2 \sin^2\beta \\
    D &= \nu_1 + \mu_2 \sin^2\beta + \nu_2 \cos^2\beta \\
    B &= (\mu_2-\nu_2)\sin\beta\cos\beta .
\end{aligned}
\end{equation}

To directly characterize the theoretical positioning accuracy of the dual-satellite solution, the two-dimensional root-mean-square positioning accuracy is defined as $\mathrm{POS}_{2D}=
\sqrt{
\operatorname{trace}
\left({\mathbf{P}}_{12,2D}
\right)
} = \sqrt{\frac{A+D}{AD-B^2}}$. 
Then, the explicit relationship between the error-ellipse intersection angle and the theoretical positioning accuracy can be expressed as
\begin{equation}
\mathrm{POS}_{2D}(\beta)
=
\sqrt{
\frac{C_0}{C_1 + C_2 \sin^2\beta}
}.
\end{equation}
where
\begin{equation}
\begin{aligned}
    C_0 &= \mu_1+\nu_1+\mu_2+\nu_2\\
    C_1 &= \mu_1\nu_1+\mu_2\nu_2+\mu_1\nu_2+\mu_2\nu_1\\
    C_2 &= (\mu_1-\nu_1)(\mu_2-\nu_2).
\end{aligned}
\end{equation}

The above expression shows that the influence of the relative angle enters the positioning accuracy only through the $\sin^2\beta$ term. Therefore, within this analytical model, the geometric complementarity becomes strongest and the fused positioning accuracy reaches its best condition as $\beta$ approaches $90^\circ$. In contrast, when $\beta$ approaches $0^\circ$, the two effective error ellipses become nearly aligned, resulting in weaker geometric complementarity and degraded positioning performance.

\section{Validation of Single-Satellite and Dual-Satellite Accuracy Prediction}
This section validates the proposed accuracy prediction framework using real Iridium Doppler measurements. The Iridium Ring Alert channel provides repeated burst transmissions with a known signal structure, enabling dense short-term Doppler observations to be obtained within a single satellite pass. These measurements are used to examine whether the predicted single-pass error ellipses are consistent with empirical positioning errors and whether the proposed dual-pass information fusion model can predict the accuracy improvement obtained by combining two passes.

\subsection{Iridium Measurement Setup and Doppler Noise Characterization}
The key acquisition parameters of the Iridium dataset are summarized in \Cref{tab:Iridiumstrategy}. Due to the burst-based transmission structure, the extracted Doppler measurements consist of short discontinuous segments. Therefore, the Doppler noise level is estimated within each burst and then averaged over the available observation period.
\begin{table}[ht]
\centering
\caption{Key parameters of Iridium signal acquisition}
\label{tab:Iridiumstrategy}
\begin{tabular}{lc}
\toprule
\textbf{Feature} & \textbf{Iridium}  \\
\midrule
Collect Point & $41.3685^\circ$, $2.1404^\circ$, 30m \\
Frequency Band & L-band (1616--1626.5 MHz) \\
Doppler Shift & High ($\pm$35 kHz) \\
Multiple Access & TDMA + FDMA \\
Transmission Type & Burst-based \\
Modulation & BPSK/QPSK ($\sim$25 kbaud) \\
Observation Duration & 23-Jan-2026 10:42:46 - 11:12:46 \\
\bottomrule
\end{tabular}
\end{table}

\begin{figure}[t]
    \centering
    \includegraphics[width=1\linewidth]{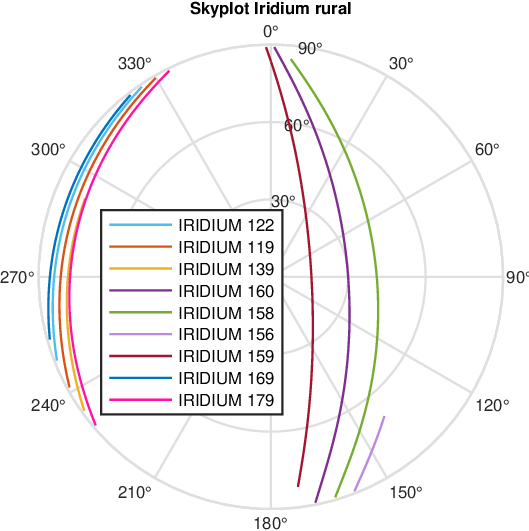}
    \caption{The skyplot of real Iridium observations}
    \label{fig:iridium_skyplot}
\end{figure}

To examine the visibility and geometric distribution of the observed satellites, the skyplot of the real Iridium observations is shown in \Cref{fig:iridium_skyplot}. Several Iridium satellites are observed during the data collection period, and their passes exhibit different azimuth and elevation characteristics. Among them, Iridium 158 and Iridium 160 are selected for detailed validation because their observation arcs are relatively complete and provide representative satellite-pass geometries.

To support the subsequent theoretical accuracy prediction, the Doppler measurement noise level needs to be estimated. Since the observations are burst-based, the noise is first estimated within each individual burst and then averaged over all available bursts. Specifically, the raw Doppler measurements are smoothed using a Savitzky–Golay filter, and the smoothed sequence is treated as the local Doppler trend. The residual between the raw Doppler sequence and the smoothed sequence is then used as the Doppler noise estimate.

\begin{figure}[t]
    \centering
    \includegraphics[width=1\linewidth]{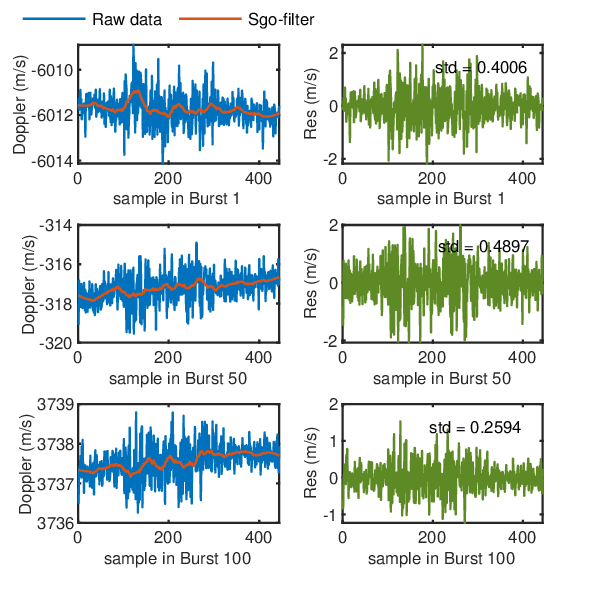}
    \caption{Representative Iridium 158 Doppler measurements and residual analysis after filtering.}
    \label{fig:Iridium_158 Noise}
\end{figure}

\begin{figure}[t]
    \centering
    \includegraphics[width=1\linewidth]{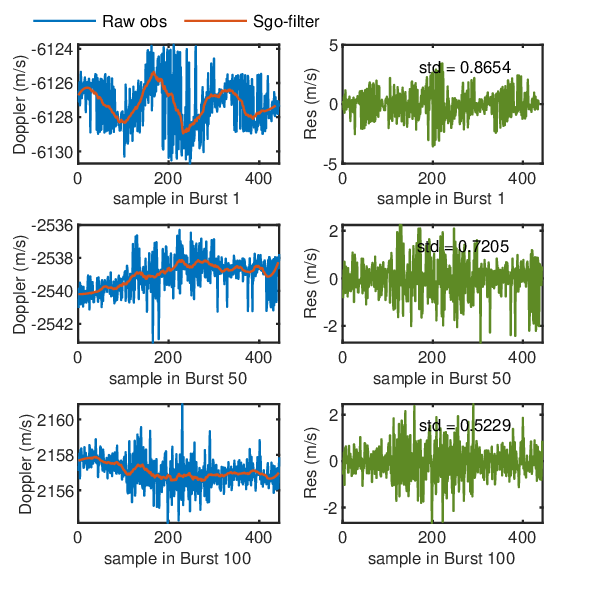}
    \caption{Representative Iridium 160 Doppler measurements and residual analysis after filtering.}
    \label{fig:Iridium_160 Noise}
\end{figure}

\Cref{fig:Iridium_158 Noise} and \Cref{fig:Iridium_160 Noise} show representative Doppler smoothing and residual results for Iridium 158 and Iridium 160, respectively. The raw Doppler observations contain visible high-frequency fluctuations, while the smoothed curves preserve the local Doppler trend. The residuals fluctuate around zero, indicating that the extracted component mainly corresponds to random Doppler measurement noise. The average Doppler noise standard deviations over the full observation period are estimated as 0.2654m/s for Iridium 158 and 0.3627m/s for Iridium 160. These values are used in the following accuracy prediction and validation.

\subsection{Single-Satellite Error-Ellipse Validation}
After estimating the Doppler noise level, the single-satellite accuracy representation is validated using real Iridium observations. For each selected satellite pass, multiple Doppler positioning trials are constructed from the available burst measurements. The receiver position and pass-specific parameters are estimated using the WLS model described in Section II, while the corresponding theoretical covariance is predicted using the single-pass information matrix derived from the Doppler observation model.

The purpose of this validation is to examine whether the predicted theoretical error ellipse is consistent with the empirical error distribution obtained from real measurements. For each pass, the positioning error samples are compared with both the empirical error ellipse and the predicted theoretical error ellipse.

\begin{figure}[t]
    \centering
    \includegraphics[width =1\linewidth]{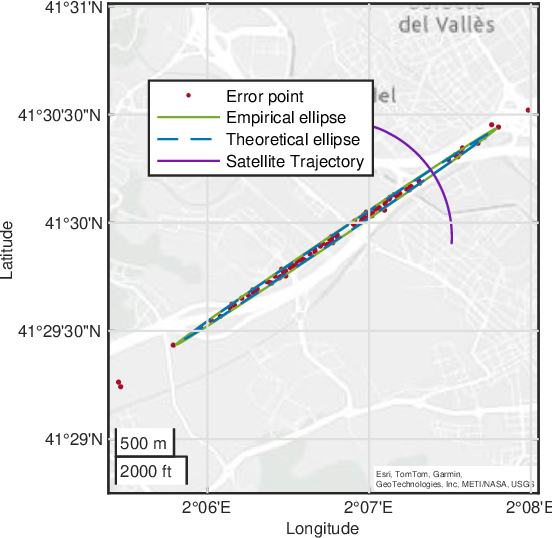}
    \caption{Verification of the Consistency Between the Experimental Error Ellipse and the Theoretical Ellipse for Iridium 158.}
    \label{fig:iridium_158 consistency}
\end{figure}

\Cref{fig:iridium_158 consistency} shows the validation result for Iridium 158. The positioning error points exhibit a clearly anisotropic distribution, which is consistent with the geometric characteristics of single-pass Doppler positioning. The empirical error ellipse and the predicted theoretical ellipse show similar dominant directions, indicating that the single-satellite accuracy representation can capture the main uncertainty direction induced by the satellite-pass geometry.

\begin{figure}[t]
    \centering
    \includegraphics[width =1\linewidth]{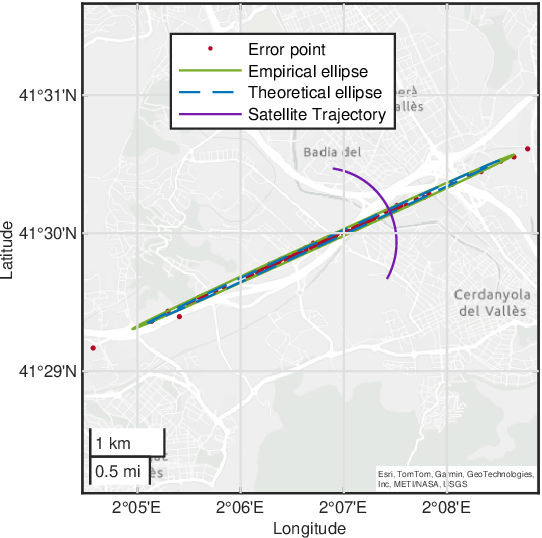}
    \caption{Verification of the Consistency Between the Experimental Error Ellipse and the Theoretical Ellipse for Iridium 160.}
    \label{fig:iridium_160 consistency}
\end{figure}

\Cref{fig:iridium_160 consistency} presents the corresponding result for Iridium 160. Although the error magnitude and ellipse orientation differ from those of Iridium 158, the theoretical ellipse remains generally consistent with the empirical error distribution. This confirms that the single-satellite method is able to predict not only the overall positioning uncertainty, but also its directional behavior under real Doppler observations.

These results validate the use of the predicted covariance as the accuracy representation of an individual Iridium satellite pass. This provides the experimental basis for the dual-satellite information fusion model evaluated in the next subsection.

\subsection{Dual-Satellite Accuracy Prediction Validation}
Based on the validated single-pass error-ellipse representation, the dual-pass accuracy prediction model is further evaluated using the joint observations from Iridium 158 and Iridium 160. In the dual-satellite solution, the two satellite passes share a common receiver position, while their clock drift and timing correction parameters are treated independently. The theoretical covariance of the fused solution is predicted using the equivalent position information matrix derived in Section II.

\begin{figure}[t]
    \centering
    \includegraphics[width =1\linewidth]{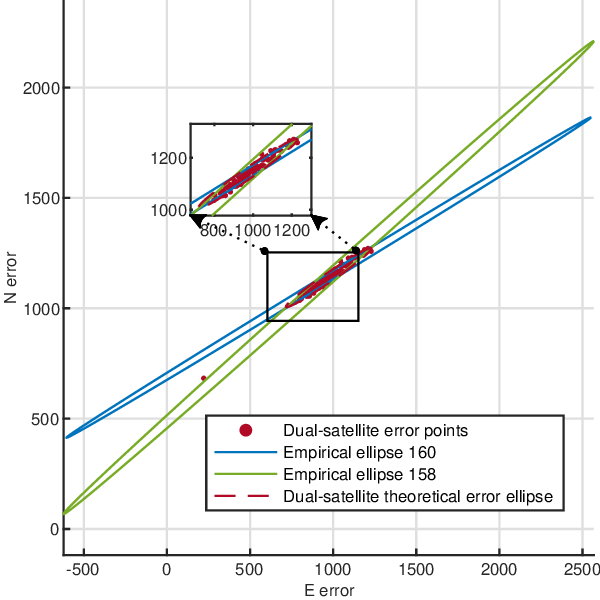}
    \caption{Comparison between the empirical error distribution and the predicted theoretical error ellipse for the fused Iridium 158 and Iridium 160 solution.}
    \label{fig:combined sats}
\end{figure}

\Cref{fig:combined sats} compares the positioning error distribution of the dual-satellite solution with the predicted theoretical error ellipses. The single-pass error ellipses provide an intuitive illustration of the directional uncertainty contributed by each satellite pass. After fusing the two passes, the positioning errors become more concentrated in the region jointly constrained by the two single-pass geometries, and the empirical spread of the dual-satellite solution is consistent with the predicted dual-satellite error ellipse. This agreement verifies that the proposed information-domain fusion model can characterize the theoretical accuracy of the dual-satellite solution. In addition, the consistency between the empirical error distribution and the predicted dual-satellite ellipse provides an indirect validation of the dual-satellite positioning implementation.

Compared with the individual-pass solutions, the dual-satellite solution exhibits a more stable and concentrated positioning error distribution. This improvement indicates that the two Iridium passes provide complementary Doppler geometry, which helps reduce the directional ambiguity inherent in single-pass positioning. More importantly, the observed improvement is consistent with the geometric information matrix derived in Section II, where the fused positioning covariance is determined by the sum of the effective position information from the two passes. Therefore, the Iridium results validate that the proposed accuracy prediction framework can characterize both the theoretical accuracy of an individual satellite pass and the accuracy improvement obtained from dual-pass fusion. This motivates the following section, where the role of pass geometry complementarity is further analyzed through the intersection-angle–accuracy relationship.

\section{Intersection-Angle and Waiting-Time Analysis}
After validating the single- and dual-pass accuracy prediction models with Iridium observations, this section investigates how the intersection-angle relationship can be used for practical geometry selection and waiting-time analysis. Long-term ORBCOMM observations are used for this purpose because they provide diverse pass-pair geometries over an extended observation period. The analysis first characterizes the quality and diversity of the observed single-pass geometries, then examines the relationship between the error-ellipse intersection angle and the achievable dual-pass positioning accuracy, and finally evaluates the waiting time required to obtain a sufficiently complementary second pass.

\subsection{ORBCOMM Measurement Setup and Pass Geometry Statistics}
The ORBCOMM signals used in this study are collected over a continuous observation period of ten days. Compared with the short-term Iridium dataset used for model validation, the long-term ORBCOMM observations provide richer pass geometries and more satellite-pair combinations. The main acquisition parameters of the ORBCOMM dataset are summarized in \Cref{tab:realdata_strategy_orbcomm}.

\begin{table}[ht]
\centering
\caption{Key parameters of Orbcomm signal acquisition}
\label{tab:realdata_strategy_orbcomm}
\begin{tabular}{lcc}
\toprule
\textbf{Feature} &  \textbf{ORBCOMM} \\
\midrule
Collect Point &  $41.5002^\circ$, $2.1129^\circ$, 130m\\
Frequency Band &  VHF (137--138 MHz) \\
Doppler Shift &  Low ($\pm$3--4 kHz) \\
Multiple Access & FDMA \\
Transmission Type & Continuous \\
Modulation  & SDPSK ($\sim$4.8 kbaud) \\
Observation Duration & 29-Mar-2026 - 09-Apr-2026 \\
\bottomrule
\end{tabular}
\end{table}

\begin{figure}[t]
    \centering
    \includegraphics[width =1\linewidth]{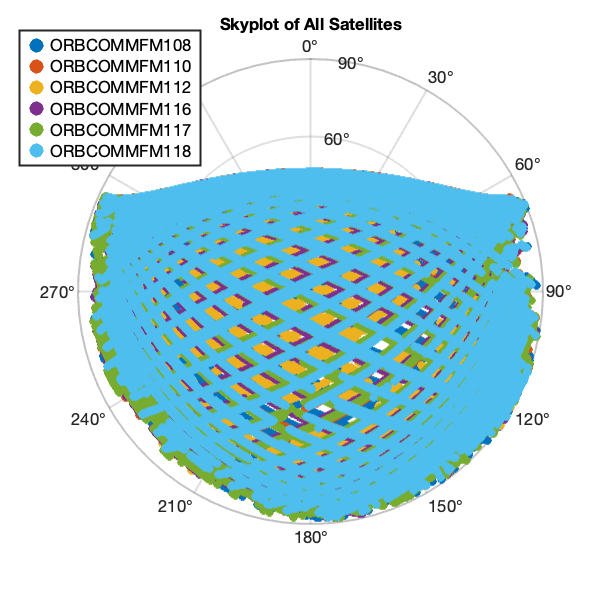}
    \caption{Skyplot of the observed ORBCOMM satellite passes during the ten-day observation period.}
    \label{fig:satplot}
\end{figure}

\Cref{fig:satplot} shows the skyplot of the observed ORBCOMM satellite passes during the ten-day observation period. The satellites cover a wide range of azimuth and elevation angles, providing diverse pass geometries for the subsequent intersection-angle analysis. For each available satellite pass, the predicted position covariance is computed using the method described in Section II, and the corresponding two-dimensional horizontal error ellipse is obtained for geometry characterization.

\begin{figure}[t]
    \centering
    \begin{subfigure}{1\linewidth}
        \centering
        \includegraphics[width=\linewidth]{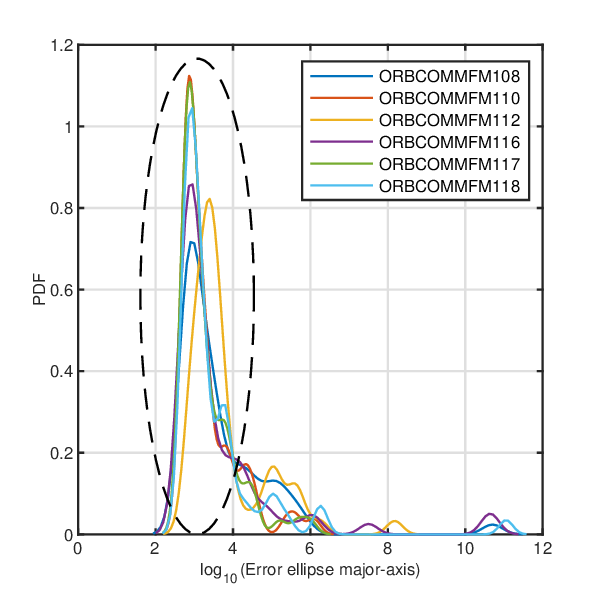} 
        \caption{}
    \end{subfigure}
    
    \begin{subfigure}{1\linewidth}
        \centering
        \includegraphics[width=\linewidth]{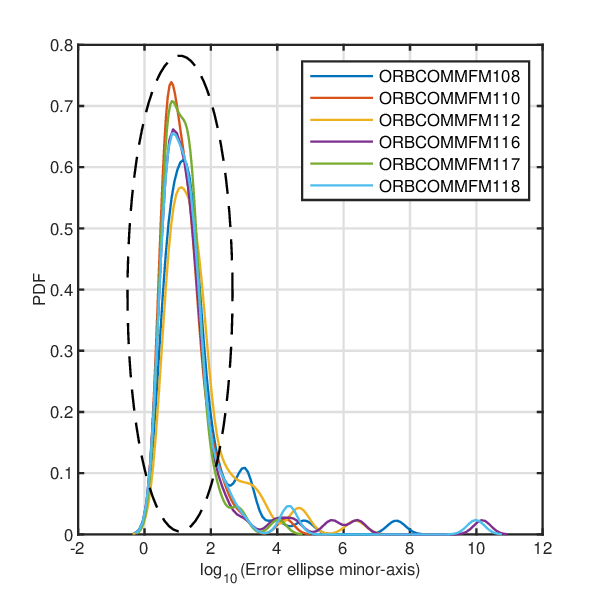}    
         \caption{}
    \end{subfigure}
  \caption{Probability density distributions of error ellipse major and minor axes derived from single-satellite positioning.}
  \label{fig:pdf}
\end{figure}

\Cref{fig:pdf} shows the probability density distributions of the major and minor axes of the predicted single-pass error ellipses derived from single-satellite positioning. The major axis represents the weakly constrained direction of a satellite pass, while the minor axis corresponds to the relatively well-constrained direction. The distributions indicate that the ORBCOMM passes have clearly nonuniform geometric quality: most passes are concentrated within a dominant accuracy range, while a small number of passes form long tails, especially in the major-axis direction. These long-tail samples correspond to weak single-pass geometries with poor observability in the dominant error direction. If included directly, they may dominate the subsequent dual-pass statistics and obscure the effect of geometric complementarity.

\begin{figure}[t]
    \centering
    \begin{subfigure}{1\linewidth}
        \centering
        \includegraphics[width=\linewidth]{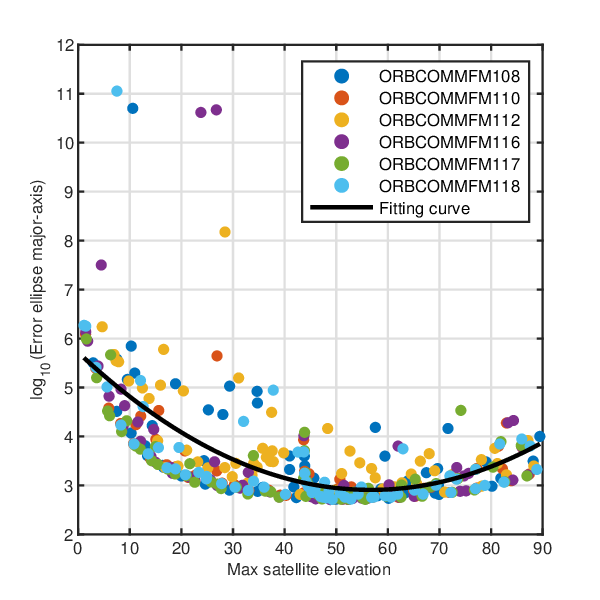} 
        \caption{}
    \end{subfigure}
    
    \begin{subfigure}{1\linewidth}
        \centering
        \includegraphics[width=\linewidth]{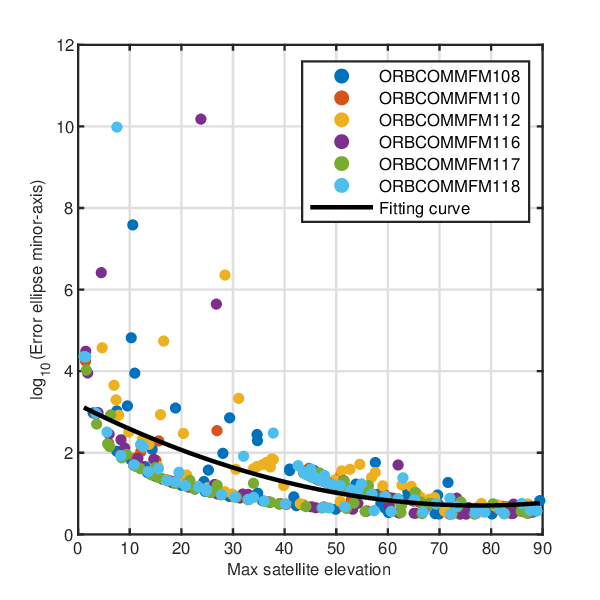}    
         \caption{}
    \end{subfigure}
  \caption{Relationship between the maximum elevation angle and the major and minor axes of the predicted single-pass error ellipses.}
  \label{fig:ele-pdf}
\end{figure}

Therefore, the error-ellipse axis distributions are used as a quality-control criterion before constructing pass pairs. Based on the main concentration regions in \Cref{fig:ele-pdf}, the upper bounds of the single-pass error-ellipse axes are set to $10^4$ m for the major axis and $10^2$ m for the minor axis. Passes exceeding either threshold are excluded from the following angle–accuracy analysis, so that the results mainly reflect the complementarity between usable satellite passes rather than extreme single-pass degeneracy.

\Cref{fig:ele-pdf} further explains the pass-quality variation by relating the error-ellipse axes to the maximum elevation angle of each pass. In general, passes with lower maximum elevation angles tend to produce larger theoretical uncertainties, which indicates weaker Doppler geometry and poorer single-pass observability. However, the major and minor axes do not vary in the same manner: the major axis is more sensitive to poor-geometry cases and shows a more pronounced degradation, while the minor axis remains comparatively more stable. This difference reflects the anisotropic nature of single-pass Doppler positioning and is consistent with our previous simulation-based conclusion \cite{liu2026geometric} that pass geometry affects the strongly and weakly constrained directions differently.

The above statistics show that the ORBCOMM dataset contains a wide range of pass geometries, including both representative and degraded single-pass configurations. Therefore, before analyzing dual-satellite geometry complementarity, it is necessary to characterize the quality of each individual pass through its predicted error ellipse. In the following subsection, available pass pairs are constructed from these single-pass statistics, and the relationship between the error-ellipse intersection angle and the achievable dual-satellite positioning accuracy is analyzed.

\subsection{Intersection-Angle–Accuracy Analysis}
Based on the pass-quality characterization in the previous subsection, this subsection analyzes how the intersection angle affects the theoretical accuracy of dual-satellite Doppler positioning. Before applying the analysis to long-term real pass pairs, a controlled geometry experiment is first conducted to isolate the effect of the intersection angle from the influence of individual pass quality. In this experiment, the major- and minor-axis lengths of the two single-pass predicted error ellipses are fixed, while only the intersection angle $\beta$ between their major axes is varied.
\begin{figure}[t]
    \centering
    \includegraphics[width =.5\textwidth]{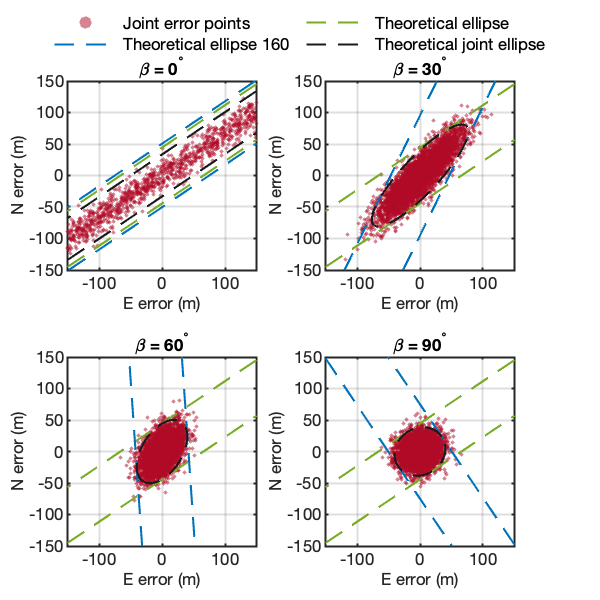}
    \caption{Joint positioning error distributions under representative intersection angles.}
    \label{fig:JointErr_TheElli}
\end{figure}
\Cref{fig:JointErr_TheElli} illustrates the joint positioning error distributions under several representative intersection angles. When $\beta = 0^\circ$, the two single-pass error ellipses are nearly aligned, and the fused error distribution remains elongated along the weakly constrained direction. As $\beta$ increases, the two passes provide more complementary geometric constraints, and the joint error distribution becomes progressively more compact. When $\beta$ approaches $90^\circ$, the dominant uncertainty directions of the two passes are nearly orthogonal, resulting in the smallest and most balanced joint error distribution.

\begin{figure}[t]
    \centering
    \includegraphics[width=\linewidth]{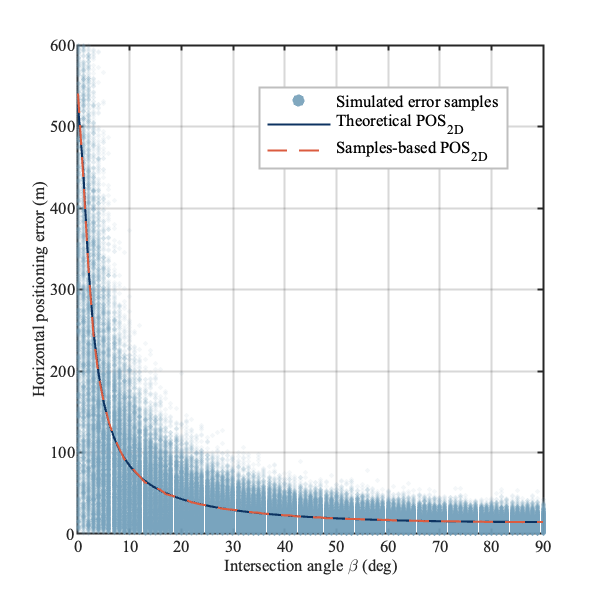}
    \caption{Comparison between theoretical $POS_{2D}$ and simulated positioning errors over different intersection angles.}
    \label{fig:formula_valid}
\end{figure}

\Cref{fig:formula_valid} further compares the theoretical $POS_{2D}$ predicted by the closed-form angle--accuracy model with the sample-based positioning results over the full range of intersection angles. For each angle, positioning error samples are generated according to the fused dual-pass covariance described in Section II. The sample-based $POS_{2D}$ is then calculated from these samples using the same definition as the theoretical metric. The theoretical curve agrees closely with the sample-based $POS_{2D}$ over the full range of intersection angles, confirming that the closed-form expression captures the dependence of dual-pass positioning accuracy on the intersection angle. Meanwhile, the simulated error distribution becomes progressively more concentrated as the angle increases, illustrating the improvement in geometric complementarity between the two satellite passes.

\begin{figure}[t]
    \centering
    \includegraphics[width=1 \linewidth]{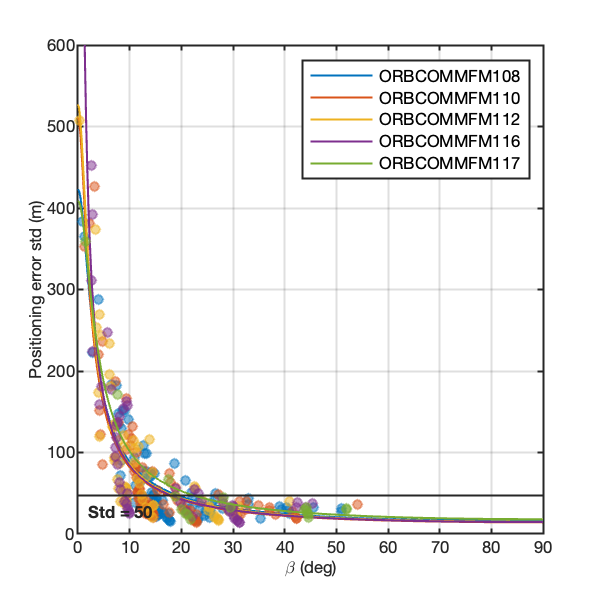}
    \caption{Relationship between the error-ellipse intersection angle and the dual-satellite theoretical positioning accuracy for different ORBCOMM satellites.}
    \label{fig:all_satellites}
\end{figure}

After the controlled analysis, the angle–accuracy relationship is further examined using real ORBCOMM pass pairs. This step evaluates whether the theoretical trend observed in the controlled experiment also appears in real long-term pass-pair geometries. \Cref{fig:all_satellites} shows the dual-satellite theoretical positioning accuracy as a function of the error-ellipse intersection angle. Although different satellites exhibit slightly different curves due to variations in their individual pass geometries, all results show a consistent decreasing trend: small intersection angles lead to poor geometric complementarity and larger positioning errors, whereas larger intersection angles significantly improve the theoretical positioning accuracy. The horizontal reference line marks the 50 m accuracy level. The results indicate that, for the analyzed ORBCOMM observations, an intersection angle of approximately $20^\circ$ is generally required to achieve this level. This provides a practical geometry-selection threshold for choosing a sufficiently complementary second pass.

These results show that the intersection angle can serve as a practical geometry-selection criterion for dual-satellite Doppler positioning. However, in real observations, a receiver cannot freely choose the second pass and must wait for a sufficiently complementary one to appear. Therefore, the following subsection analyzes the relationship between waiting time, achievable intersection angle, and theoretical positioning accuracy.

\subsection{Waiting Time for Complementary Pass Acquisition}
After the intersection-angle–accuracy relationship is established, a practical question naturally arises: after one usable satellite pass has been observed, how long does the receiver need to wait for a sufficiently complementary second pass? This question is important for sparse LEO signal-of-opportunity positioning, where the receiver cannot arbitrarily select satellite geometry but can only use the passes that become available over time.

To investigate this issue, the long-term ORBCOMM observations are analyzed from a waiting-time perspective. For each maximum waiting time, all candidate pass pairs whose time separation does not exceed this limit are considered, and the pair with the minimum theoretical $\mathrm{POS}_{2D}$ is selected. This produces a best-achievable accuracy curve as a function of maximum waiting time.

\begin{figure}[t]
    \centering
    \includegraphics[width =1\linewidth]{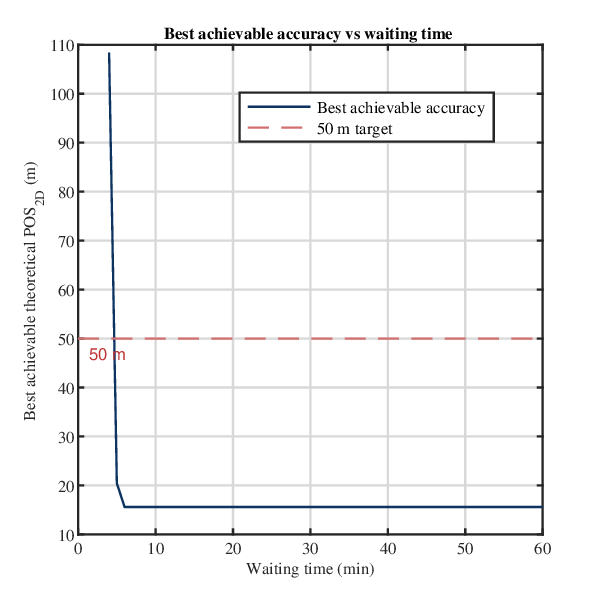}
    \caption{Best achievable theoretical positioning accuracy as a function of maximum waiting time.}
    \label{fig:intersection}
\end{figure}

\Cref{fig:intersection} shows the best achievable theoretical positioning accuracy under different maximum waiting-time limits. When only very short waiting times are allowed, the best achievable accuracy is still limited by the geometry of the available pass pairs. As the maximum waiting time increases, more candidate second passes become available, and the best achievable accuracy improves rapidly once a sufficiently complementary pass pair appears.

For the analyzed ORBCOMM observations, the best achievable theoretical accuracy drops below the 50 m target within approximately several minutes of maximum waiting time. After this point, the curve further decreases to about 15 m and then remains almost unchanged up to 60 min. This indicates that, in this dataset, a highly complementary second pass becomes available within a short waiting period, and extending the waiting time beyond this point does not necessarily provide additional improvement. Therefore, the waiting-time result should be interpreted as a best-case availability curve: the achievable accuracy is governed not only by how long the receiver waits, but also by whether a geometrically complementary pass pair appears within the allowed time window.

\section{Conclusion}
This paper investigated theoretical accuracy prediction and geometry selection for dual-pass LEO Doppler positioning in sparse signals-of-opportunity scenarios. A single-pass accuracy representation was first validated using real Iridium Doppler measurements. The results showed that the predicted error ellipses are generally consistent with the empirical positioning error distributions, confirming that the single-pass covariance representation can capture both the magnitude and the dominant direction of positioning uncertainty.

Based on this validated single-pass representation, an information-domain dual-pass accuracy prediction model was developed. By enforcing a common receiver position while treating clock drift and timing correction parameters as pass-specific unknowns, the effective geometric information from two satellite passes was combined. Real Iridium observations further showed that the dual-pass solution provides more stable and concentrated positioning performance than individual-pass solutions, demonstrating the effectiveness of the proposed information fusion model.

To characterize which dual-pass configurations are more effective, this paper derived an analytical relationship between the error-ellipse intersection angle and the theoretical dual-pass positioning accuracy. The intersection angle was defined as the angle between the major axes of two single-pass predicted error ellipses. The derived expression shows that larger intersection angles provide stronger geometric complementarity and lower theoretical positioning error. Controlled simulations further verified that the closed-form angle--accuracy expression is consistent with sample-based positioning results.

Long-term ORBCOMM observations were then used to evaluate the practical significance of the proposed intersection-angle model. The pass-pair analysis confirmed that larger intersection angles generally lead to lower theoretical positioning errors, and an intersection angle of approximately $20^\circ$ is required to achieve a theoretical positioning accuracy of about 50m in the analyzed dataset. The waiting-time analysis further showed that, after one usable pass has been observed, the best achievable theoretical accuracy can drop below the 50m target within a short waiting period and then remains nearly unchanged as the maximum waiting time further increases. This indicates that the achievable accuracy is governed not only by waiting time, but also by whether a geometrically complementary second pass appears within the allowed time window.

Overall, the proposed framework provides a quantitative link between individual-pass geometry, dual-pass information fusion, intersection-angle-based accuracy prediction, and temporal availability. The results demonstrate that sparse LEO Doppler positioning should not rely solely on satellite availability, but should also account for the geometric complementarity between passes. The derived angle--accuracy relationship and waiting-time analysis can support geometry-aware satellite-pair selection, observation scheduling, and accuracy availability assessment for sparse LEO signals-of-opportunity PNT applications.

\bibliographystyle{IEEEtran}
\bibliography{ref}

\end{document}